# Multi-wavelength study of a new Galactic SNR G332.5-5.6

M. Stupar,[1*] Q. A. Parker,[1,2] M. D. Filipović,[3] D. J. Frew,[1] I. Bojičić,[1]
and B. Aschenbach[4]

[1] *Department of Physics, Macquarie University, Sydney, NSW 2109, Australia*
[2] *Anglo-Australian Observatory, P.O. Box 296, Epping, NSW 1710, Australia*
[3] *University of Western Sydney, Locked Bag 1797, Penrith South DC, NSW 1797 Australia*
[4] *Max-Planck-Institut für extraterrestrische Physik, Giessenbachstrasse, D-85741 Garching, Germany*

**ABSTRACT**

We present compelling evidence for confirmation of a Galactic supernova remnant (SNR) candidate, G332.5-5.6, based initially on identification of new, filamentary, optical emission line nebulosity seen in the arcsecond resolution images from the AAO/UKST Hα survey. The extant radio observations and X-ray data which we have independently re-reduced, together with new optical spectroscopy of the large-scale fragmented nebulosity, confirms the identification. Optical spectra, taken across five different, widely separated nebula regions of the remnant as seen on the Hα images, show average ratios of [NII]/ Hα = 2.42, [SII]/ Hα = 2.10, and [SII] 6717/6731 = 1.23, as well as strong [OI] 6300, 6364Å and [OII] 3727Å emission. These ratios are firmly within those typical of SNRs.

Here, we also present the radio-continuum detection of the SNR at 20/13cm from observations with the Australia Telescope Compact Array (ATCA). Radio emission is also seen at 4850 MHz, in the PMN survey (Griffith & Wright 1993) and at 843 MHz from the SUMSS survey (Bock, Large & Sadler 1999). We estimate an angular diameter of ~30' and obtain an average radio spectral index of α=-0.6 ± 0.1 which indicates the non-thermal nature of G332.5-5.6. Fresh analysis of existing ROSAT X-ray data in the vicinity also confirms the existence of the SNR.

The distance to G332.5-5.6 has been independently estimated by Reynoso & Green (2007) as 3.4 kpc based on measurements of the HI λ21 cm line seen in absorption against the continuum emission. Our cruder estimates via assumptions on the height of the dust layer (3.1 kpc) and using the Σ-D relation (4 kpc) are in good agreement.

## 1 INTRODUCTION

Supernovae are a principal source of energy injection into the interstellar medium (ISM). Multiwavelength studies of supernova remnants (SNRs) provide crucial information on several key aspects of stellar evolution and the physical behavior of shocked material. SNRs may contribute up to 50% of the emission measure observed for the Galactic warm ionised medium (Slavin 2000). Supernova explosions also enrich the ISM with heavy elements synthesised in their high mass progenitor stars, while ejected material and the attendant shock wave shape and heat the ISM. They influence Galactic magnetic fields (Milne 1990) whilst the shock waves may trigger new bursts of star formation and accelerate energetic cosmic rays observed throughout the Galaxy. SNRs can reveal Galactic abundance gradients by virtue of the swept up local ISM, and yield information on the death rate of high mass stars. They significantly impact on the morphology, kinematics and ionization balance of the Galaxy.

An up-to-date inventory of Galactic SNRs is an important aid in our understanding of Galactic star formation history. However, the total number known (currently 265 according to the continuously updated, on-line compilation curated by Green 2006) compared to those expected (~1000, Case & Bhattacharya 1998) implies a significant fraction of Galactic remnants remains to be uncovered. This deficit is thought to be due to selection effects mitigating against detection of both the youngest, small scale remnants and the oldest remnants of large angular size which may be highly fractured and dissipating into the ambient ISM. Also, from all currently known Galactic SNRs there are only 15 with |b|≥5° but their distribution is asymmetrical. Only 4 of them are in the Southern Hemisphere, while the rest are above + 5° (Green 2006). Although this asymmetry may be related to Gould's belt (Stothers & Frogel 1974), there is no evidence of the same asymmetry in low Galactic latitude SNRs, which are not statistically different. SNRs at high Galactic latitudes are also very rare. Those that are found are ideal probes for a variety of studies such as abundance variations, environmental influence etc.

---

[*] E-mail: mstupar@physics.mq.edu.au

The advent of the AAO/UKST Hα survey (Parker et al. 2005) with its combination of 5 Rayleigh sensitivity, arcsecond resolution, and 4000 sq. degree coverage, offers a fresh opportunity to improve this situation from a non-radio perspective. This survey has already provided significant discoveries of planetary nebulae (e.g. Parker et al. 2006), HII regions and the uncovering of other filamentary SNR candidates from a previous small-scale study (Walker, Zealey & Parker 2001).

However, we have undertaken the first large-scale, systematic search for new Galactic SNRs over the entire survey area. Candidates were selected based initially on detection of new, coherent, optical structures identified from the Hα emission. Both small- and large-scale, low-surface brightness nebulosities were mapped. Results for 21 new, confirmed Galactic SNRs are presented in Stupar (2007) while Stupar, Parker & Filipović (2007) presents a detailed study of another confirmed new remnant G315.1+2.7. In this paper we report the discovery of the first confirmed Galactic SNR to be found from the Hα survey: designated G332.5-5.6. We combine the new narrow-band optical imagery and emission line spectroscopy with fresh reductions of the extant radio data at four frequencies (843, 1384, 2432 and 4850 MHz) and with existing ROSAT X-ray observations to present a compelling picture to confirm this interesting new SNR.

## 2 IDENTIFICATION OF THE NEW GALACTIC SNR: G332.5-5.6

As part of a search for new planetary nebulae (PNe) from the AAO/UKST Hα survey (e.g. Parker et al. 2005) a new Galactic supernova remnant (SNR) was uncovered. The most prominent optical component of this remnant was first identified from the original Hα survey films in 1998 and due to its unusual morphology, dubbed the `paperclip'. The Hα image of the paperclip, as taken from the online digital survey (SuperCOSMOS Halpha Survey - SHS hereafter) data[1] is shown on the top left hand corner of the panel of the most significant optical components of this SNR given in Fig. 4 (see later). The instinctive, loopy morphology is clearly unlike that of a planetary nebula or, due to lack of obvious dust or proximity to star forming regions, an HII region.

Nevertheless, as part of an extensive programme of complete follow-up optical spectroscopy of newly identified PNe candidates, QAP & DJF obtained a spectrum of this object using the South African Astronomical Observatory (SAAO) 1.9-m telescope in February 2004 (see section 2.3). The spectrum immediately indicated a potential SNR based on the great strength of the [SII] 6717/6731 lines relative to Hα, together with the very strong [OII] 3727Å line and prominent [OI] 6300 and 6364Å lines. The preliminary wavelength and flux calibrated 1-D spectrum is shown in Fig. 1 obtained from summing the brightest 20 pixels of the extended nebula along the slit. The 2-D sky-subtracted spectrum of the paperclip is also shown in Fig. 2 to indicate the stratification and variation in line strengths across the filamentary nebula.

Next a search for possible connection of the nebulosity with existing radio data was performed using SIMBAD which revealed a radio source in close proximity to the paperclip: PMN J1642-5429, which comes from the Parkes-MIT-NRAO (PMN) survey (Griffith & Wright 1993). Increasing the search radius to 1-degree revealed a total of 11 apparently discrete PMN sources, with 8 of them arranged in a fractured ring pattern with a diameter of ~30'.

A search of the up-to-date, on-line Green catalogue of all currently known Galactic SNRs gave no matches with known SNRs within an 1-degree search radius, nor was a specific entry found in his supplementary list of possible and probable Galactic SNRs. Further investigation revealed that, in a study of SNR candidates from the Parkes 2400 MHz survey by Duncan et al. (1997), an SNR candidate G332.5-5.6, with a diameter of ~ 30', was suggested in this region based on additional PMN radio data at 4850 MHz among 24 other candidates. It has three parallel radio 'filaments' seen at 4850 MHz but was not covered at 2400 MHz. Though G332.5-5.6 did not show typical SNR morphology, Duncan et al. (1997) concluded that it is a likely new SNR. Unfortunately, their work did not comprise measurements of the radio flux of most of their SNR candidates with which to confirm their nature. Hence, G332.5-5.6 has never appeared in the list of known Galactic SNRs curated by Green. Later, independent optical discovery of the paperclip component by Parker, Frew & Stupar (2004) and a preliminary investigation based on optical imaging and spectroscopy strongly indicated the likely SNR nature of the emission.

With the optical spectrum of the paperclip indicating shocked-gas and with the radio detections revealing the possibility of a coherent, though fractured, oval structure, we returned to the Hα data to look for other optical nebulosities that might be associated. Several 30' × 30' regions in Hα with the matching broad-band short-red (SR) data were extracted from the SHS in the vicinity of the paperclip. Quotient images (Hα and SR image division) were obtained which revealed faint nebulosity otherwise masked by the rich stellar field. This is an effective technique as the point-spread functions of the Hα and SR bands are well matched (Parker et al. 2005). The other filamentary structures found in the region not only exhibited very similar position angles to the existing radio maps but were also positionally coincident with three of the earlier PMN radio source detections: PMN J1642-5436, PMN J1643-5438 and PMN J1643-5432a. The ~1° Hα/SR quotient map is given in Fig. 3 which reveals these filamentary emission structures. A SIMBAD search also revealed three ROSAT X-ray sources within ~ 30' of the centre of G332.5-5.6. One of these sources, 1RXS J164251.3-542955, is resolved and sits completely within the central bar-like filament of the remnant seen at all four observed radio frequencies (see section 3).

---

[1] http://www-wfau.roe.ac.uk/sss/halpha/

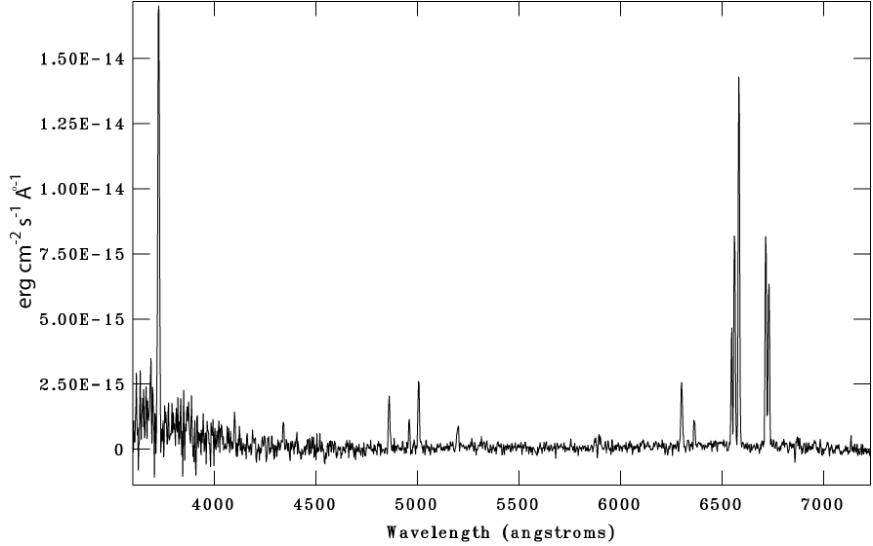

**Figure 1.** The low-dispersion, flux calibrated optical spectrum taken across the central brightest region of the paperclip nebula (see the top left panel of Figure 4). Note the strong, indicative SNR lines of [OII] 3727Å in the far blue and the very strong [SII] lines relative to Hα in the red which is a clear sign of shocked gas.

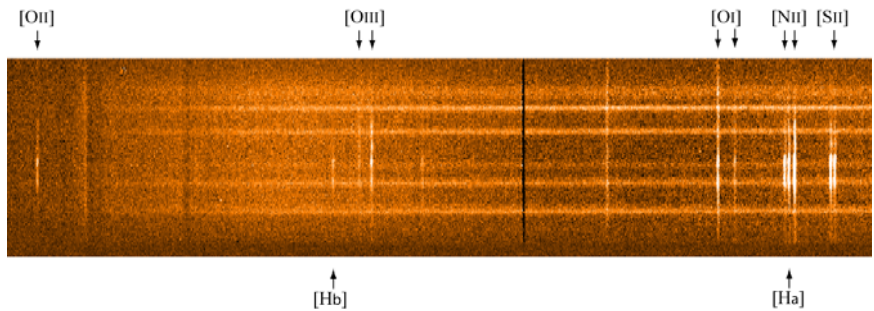

**Figure 2**. 2-D sky-subtracted spectral image of the paperclip. This clearly reveals the stratification and variable line strength across the key diagnostic emission lines. All major lines which identify an SNR are clearly seen, from [OII] 3727Å on the left side of image to the [SII] 6717 and 6731Å at right. The vertical line near centre is the oversubtracted strong night-sky line at 5577Å. It was this preliminary spectrum and the relative [SII] line strength compared to Hα which first led us to identify the paperclip as part of an SNR.

### 2.1 Follow-up optical spectroscopy

As part of a project to search for and follow-up new Galactic SNR candidates identified from the AAO/UKST Hα survey (Stupar 2007, and Stupar et al., in preparation) QAP and MS obtained further SAAO optical spectroscopy in July 2004 for five of the additional suspected optical components of G332.5-5.6 identified in the area from the quotient map in Fig. 3. An accurate World Coordinate System (WCS) incorporated in the extracted FITS image data from the SHS allows accurate positions of the new filaments to be obtained and used to direct optical follow-up. Table 1 summarises our spectroscopic observations for G332.5-5.6.

### 2.2 Low resolution optical spectra

The paperclip nebulae (size ~4.5'), discovered by us in 1998, is only the most prominent optical component of the SNR revealed by the SHS survey at RA = $16^h42^m17^s$ and δ = -54°28'33" (J2000). The identification spectrum of the paperclip was obtained on 16th February 2004 with the CCD spectrograph on the 1.9-m telescope of the South African Astronomical Observatory (SAAO). The night was photometric and the exposure time was 1200 seconds. The low dispersion grating (G7; 300 lines/mm) was used, and the slit subtended 100 × 24 arcsecs on the sky, fixed east-west. The spectral resolution was ~ 7Å and covered 3500-7300Å. Flux calibration was via observations of spectrophotometric standard star LTT4364 taken on the same night. Standard IRAF slit-spectra reduction procedures were used to produce the flux-calibrated spectrum in Fig. 1. The observed line strengths and relevant ratios of the [OII], [OI], [NII] and [SII] lines allowed straightforward classification as an SNR according to the diagnostic regime of Fesen, Blair & Kirshner (1985). The flux ratios of the main spectral lines of [SII] 6717/6731 = 1.30, [NII]/Hα = 2.41 and [SII]/Hα = 2.00 are typical of SNRs. The very strong [OII] line at 3727Å and the observed ratio of [SII]/Hα is quite different to the values associated with HII regions or PNe (e.g. [S II]/Hα ≤ 0.4).

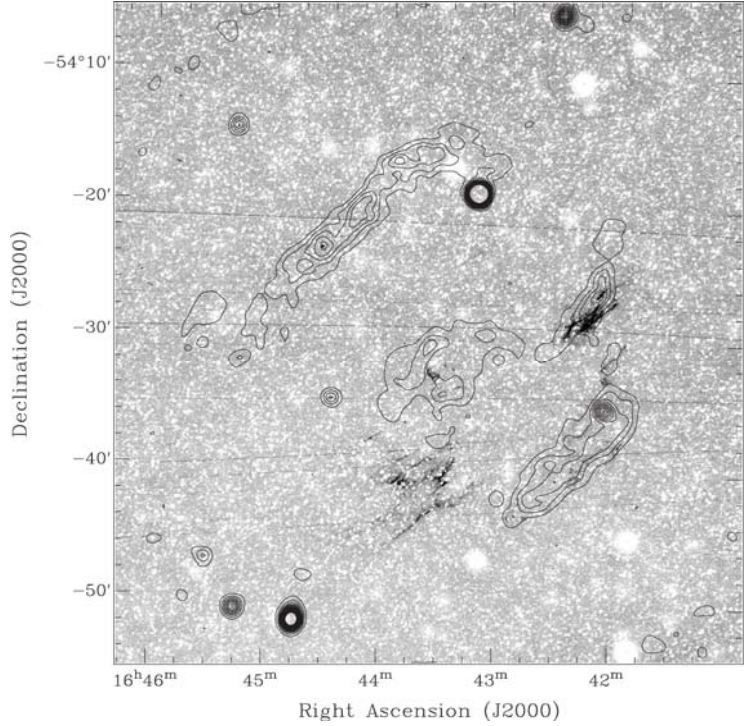

**Figure 3**. 1°×1° difference map centred at RA = 16$^h$43$^m$02$^s$ and Dec = -54°34'51" (J2000) between the Hα survey data and its matching broad-band SR counterpart obtained from the on-line SuperCOSMOS Hα survey together with the overlaid axes grid. This clearly reveals the extensive, fractured but coherent filamentary nebulosities associated with this new Galactic SNR. The faint, approximately horizontal streaks are photographic artifacts. Radio contours from the 13cm ATCA data of 2, 3, 5, 7, 10, 15, 20, 50 and 100 mJy beam$^{-1}$ are also overlaid so that the association between the optical and radio components can be clearly seen.

The same telescope and spectrograph and low resolution grating was used to obtain additional spectra of two of the other faint, optical filamentary structures of G332.5-5.6 on July 20, 2004, labeled b) and c) in Fig. 4 which gives fine detail close-ups of the fragmented optical structures of the remnant from the SHS data so that the locations of the spectroscopic slits can be seen. Unfortunately, a poor photometric night did not allow proper estimate of the Balmer decrement for these components nor a decent flux calibration. The line ratios from adjacent close lines should not be affected and these ratios are in agreement between the low resolution spectra taken across all 3 components. In Table 2 we present observed line ratios (fluxes) relative to Hβ for the low resolution spectra from the initial January 2004 photometric night and the subsequent non-photometric July 2004 run where a flux calibration was still attempted using observations of photometric standard star LTT 7379. The line intensities, corrected for extinction for the photometric January observations, are also given. Figure 5 shows a 1-D, low resolution, calibrated spectra of G332.5-5.6 across filament b) in Fig. 4 which exhibits the same spectral characteristics as in Fig. 1. The estimated line flux ratios for components b) and c) are: [SII] 6717/6731= 1.31 and 1.39; [N II]/Hα= 3.25 and 1.73; [SII]/Hα= 2.32 and 1.84. The observed, relative line strengths and ratios across these different, widely separated optical filaments, confirm that they form part of the same new Galactic SNR.

**Table 1.** Spectral observation log for the various optical components of G332.5-5.6. The SAAO 1.9m telescope with the CCD spectrograph and 1200 sec exposure times was used for all the low and medium dispersion observations.

| Date | Figure 4 Image | Grating$^a$ | Slit R.A. (h m s) | Slit Dec. (° ' ") |
|---|---|---|---|---|
| 16/02/2004 | a) | G7 | 16 42 17 | −54 28 33 |
| 20/07/2004 | b) | G7 | 16 43 53 | −54 41 02 |
| 20/07/2004 | c) | G7 | 16 43 34 | −54 43 14 |
| 25/07/2004 | e) | G5 | 16 42 24 | −54 29 26 |
| 26/07/2004 | c) | G5 | 16 43 34 | −54 43 14 |
| 26/07/2004 | d) | G5 | 16 43 33 | −54 32 57 |

$^a$ G7 Grating=300 lines/mm covering 3400–7500Å; G5 Grating=1200 lines/mm covering 6150–6900Å

### 2.3 Medium resolution spectra

The SAAO spectra for components c), d) and e) were obtained on July 25 and 26, 2004 with a medium resolution 1200 lines/mm grating (G5). In combination with a 2.5 arcsec slit, G5 gave an instrumental resolution of ~1.5Å over the region 6150 to 6900Å. With this red spectrum, line ratios relative to Hβ were not possible. Table 3 hence gives the flux ratio relative to Hα=100 for the most prominent lines calibrated according to observations of photometric standard stars LTT7379 and LTT1020.

The [SII] 6717/6731 line ratios between the high and low resolution spectra agree as does the average ratio of [N II]/Hα. However, a ~15% variation is seen for the [SII]/Hα ratio. This may reflect a small offset in the slit position between the different exposures. Figure 6 presents the 1-D spectra obtained with the medium resolution grating around Hα for slit position d). It is well known that different parts of an SNR can give rise to significant differences in some observed line ratios (see Fesen, Blair & Kirshner 1985).

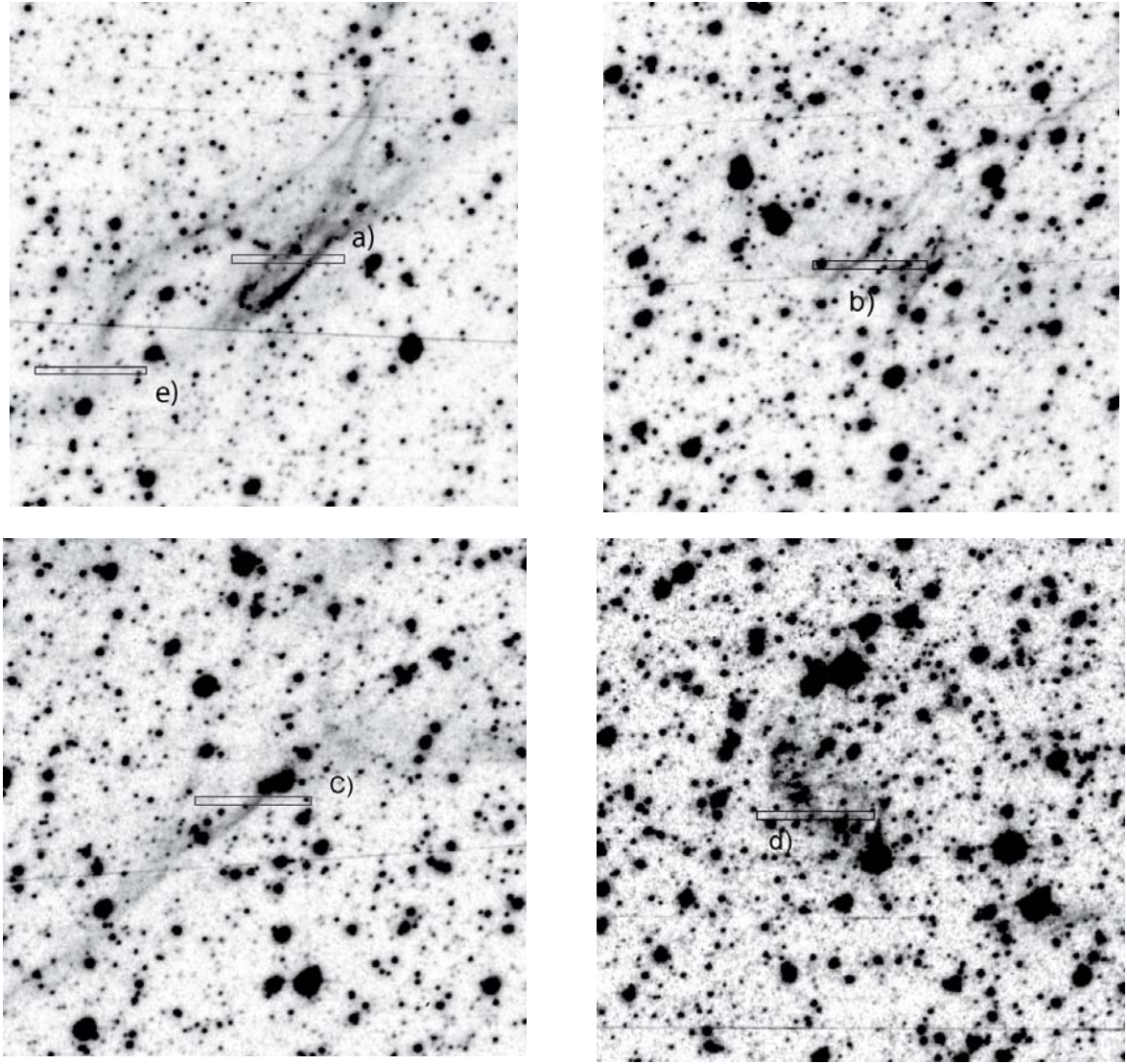

**Figure 4**. AAO/UKST Hα 4' × 4' survey images of the four most prominent optical filaments of the new SNR. The top left panel shows the bright `paperclip' nebulosity. The spectrum is shown in Fig. 1. The approximate slit positions for the spectral observations of these filaments are also indicated as black rectangles. Precise slit positions are in Table 1.

## 3 RADIO OBSERVATIONS

### 3.1 ATCA radio observations at 1384 MHz (20 cm) and 2368 MHz (13 cm)

As part of our investigation we trawled existing archives of radio data and discovered that radio-continuum observations were made in this region by R. Duncan (ATCA project C 505, September 1996) with the ATCA 375-m configuration as the part of an unpublished, high-resolution survey of new and large SNR candidates in the Southern Galactic Plane first identified in the PMN (Duncan et al. 1997). The total observing time was 8.61 hours with ~ 3.5 hours on this particular source. The source was observed at both 13 cm and 20 cm wavelengths with a 128 MHz bandwidth centered on 2368 and 1384 MHz, respectively. For all observations the primary calibrator PKS 1934-638 (with assumed flux of 14.9 Jy at 1384 MHz and 11.1 Jy at 2368 MHz) was used for bandpass and absolute flux calibration. The radio source PKS 1722-55 was used as a secondary calibrator for antenna gains and instrumental polarization calibration.

Because of the large ~ 30' angular size of SNR 332.5-5.6 compared to the beam size, the only efficient option to map the source fully was mosaicing. Duncan, a highly experienced radio observer, chose to observe the whole object using 21 different mosaic pointings with 19 of them spaced on a hexagonal ~ 11' grid (i.e. half the primary beam size at 13 cm) in

**Table 2.** Line fluxes relative to Hβ=100 for three filaments in the SNR G332.5-5.6 from spectra obtained with the 300 lines/mm grating (G7). Extinction was estimated for the January 2004 observations only (see Section 6) and therefore I(λ) is for this set of observations. The other two columns with F(λ) are for July 20, 2004 observations. Extinction was calculated for the best photometric night on January 16, 2004.

| λ (Å) | Line | Flux (Hβ=100) | | | |
|---|---|---|---|---|---|
| | | $F(\lambda)^a$ | $I(\lambda)$ | $F(\lambda)^b$ | $F(\lambda)^c$ |
| 3727 | [O II] | 925 | 1164 | 1344 | 892 |
| 4101 | Hδ | 46 | 54 | – | 35 |
| 4340 | Hγ | 45 | 50 | – | – |
| 4861 | Hβ | 100 | 100 | 100 | 100 |
| 4959 | [O III] | 49 | 48 | 108 | 49 |
| 5007 | [O III] | 130 | 126 | 278 | 168 |
| 5199 | [N I] | 58 | 55 | – | 223 |
| 5755 | [N II] | 9 | 8 | – | 13 |
| 5876 | [He I] | 23 | 19 | – | – |
| 6300 | [O I] | 140 | 109 | 253 | 728 |
| 6364 | [O I] | 48 | 37 | – | – |
| 6548 | [N II] | 216 | 162 | 994 | 930 |
| 6563 | Hα | 381 | 286 | 1327 | 2468 |
| 6583 | [N II] | 640 | 479 | 3320 | 3354 |
| 6717 | [S II] | 380 | 280 | 1744 | 2645 |
| 6731 | [S II] | 314 | 231 | 1330 | 1903 |
| 7136 | [Ar III] | 14 | 10 | – | 79 |
| 7325 | [O II] | 51 | 35 | – | – |
| Extinction | | | | | |
| c | | 0.39 | | | |
| E(B-V) | | 0.27 | | | |
| $A_v$ | | 0.83 | | | |
| Line Ratios | | | | | |
| [S II] 6717/6731 | | 1.21 | | 1.31 | 1.39 |
| [N II]/Hα | | 2.25 | | 3.25 | 1.73 |
| [S II]/Hα | | 1.82 | | 2.32 | 1.84 |

Hβ flux: Slit pos. a=1.86×10$^{-14}$; Slit pos. b=1.83×10$^{-15}$; Slit pos. c=7.90×10$^{-16}$ in units of ergs cm$^{-2}$ s$^{-1}$.
Measured fluxes for the brightest lines are accurate to ±5% based on repeat observations.

**Table 3.** Ratio of the optical emission lines for G332.5-5.6 from the medium resolution grating (1200 lines/mm; G5) relative to Hα=100. First flux column is from July 25 while second and third flux columns are from July 26, 2004 observations. For radial velocities estimates, the brightest lines of [NII] 6548 and 6584Å, Hα and [SII] 6717 and 6731Å were used.

| λ (Å) | Line | Flux$^a$ | Flux$^b$ | Flux$^c$ |
|---|---|---|---|---|
| 6300 | [O I] | 28 | 43 | 102 |
| 6364 | [O I] | 8 | 13 | 39 |
| 6548 | [N II] | 44 | 38 | 88 |
| 6563 | Hα | 100 | 100 | 100 |
| 6583 | [N II] | 148 | 143 | 266 |
| 6717 | [S II] | 99 | 81 | 201 |
| 6731 | [S II] | 90 | 82 | 144 |
| Line Ratios | | | | |
| [S II] 6717/6731 | | 1.09 | 0.99 | 1.39 |
| [N II]/Hα | | 1.92 | 1.81 | 3.54 |
| [S II]/Hα | | 1.89 | 1.63 | 3.45 |
| Radial velocities (km s$^{-1}$) | | -40±4.7 | -11.3±9.8 | -11.3±9.8 |

Hα flux: Slit pos. a=6.90×10$^{-15}$; Slit pos. b=3.62×10$^{-15}$; Slit pos. c=1.11×10$^{-15}$ in units of ergs cm$^{-2}$ s$^{-1}$.
Fluxes for the brightest lines are accurate to ±10%

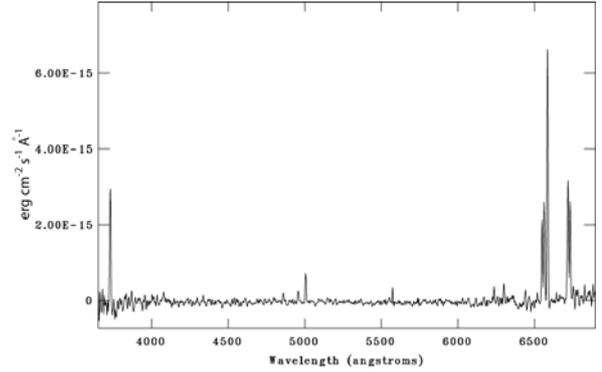

Figure 5. 1-D calibrated low resolution spectrum of G332.5-5.6 from July 20, 2004 for slit position b) in Fig. 4. Spectral coverage is between 3600 and 7000Å. Strong lines characterizing supernova remnants, are clearly seen.

such a way as to give the best, uniform coverage of the whole area (Fig. 7). The remaining two pointings were cleverly centered directly on two very strong, compact radio objects (potentially strong confusing sources) located in the vicinity of the SNR candidate. We extracted all these data from the public archive and undertook a completely fresh analysis (none appears to exist elsewhere in the literature despite extensive searches). The data reduction was carried out with the MIRIAD software package (Sault & Killen 2004). After editing and calibration were performed 'dirty' images were produced using multi-frequency synthesis and uniform weighting. The noise level across the mosaic image varies because of the low sensitivity at the edges of the mosaic pointings. However, close to bright sources, the confusion noise is much greater then the thermal system noise.

As the field contains both small and large scale structures, neither of the two mosaicing deconvolution tasks MOSMEM and MOSSDI can be successfully applied. Hence, deconvolution of the mosaic dirty images, was done with the following procedure: first, the MOSSDI task, which uses a Steer, Dewdney & Ito (SDI) CLEAN algorithm and which can generally be better for images containing point sources (Sault & Killen 2004), was applied to the dirty maps. The maximum-entropy based deconvolution task MOSMEM was then run over the residual images. Secondly, the MOSSDI clean component image was convolved with a model by using a Gaussian beam. Finally, both images were added together to produce final maps and restored with circular beam sizes. The same procedure was applied to all four Stokes parameters. The radio imaging parameters are presented in Table 4.

In Figs. 8 and 9 we show the newly processed ATCA images at 13 and 20 cm. In all the radio images available, SNR G332.5-5.6 consistently appears as three broadly parallel bands going from the SE-NW (positional angles ~-45°). There appears to be a connection between the southern and middle structures at the western edge and a small tongue of radio emission from the middle structure pointing towards the western edge of the northern filament which is particularly evident in the 20cm ATCA image of Fig. 9. Each of the 3 semi-discrete, parallel components exhibits internal substructures on closer scrutiny. This is particularly evident in the northern filament of the 13 cm image in Fig. 8 with a chain of internal clumps. The regions of apparently negative flux around and inside the SNR G332.5-5.6 (so called 'negative bowl') indicates that some of this SNRs large scale structure is filtered-out from our maps so the data needs to be interpreted carefully. Common when dealing with large-scale radio structures the sidelobes and negative artifacts could not be effectively removed in the cleaning process because of the missing short-spacing flux and incomplete *uv* coverage.

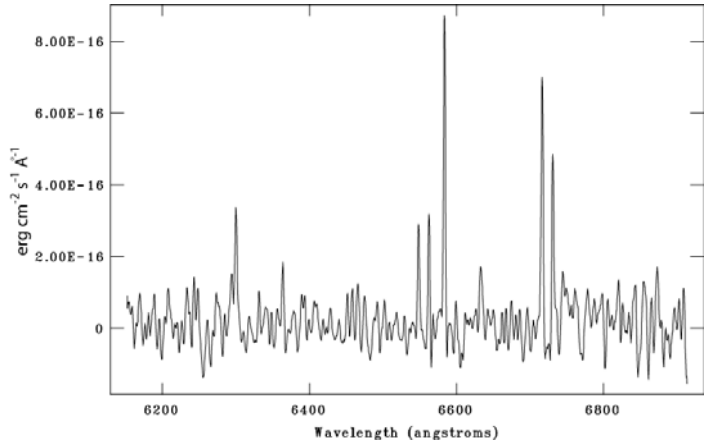

**Figure 6**. 1-D flux calibrated, medium resolution spectrum of G332.5-5.6 from July 26 2004, position d) on Figure 4, from 6150 to 6900Å. All prominent lines are clearly seen: [OI] at 6300 and 6364Å [NII] at 6548 and 6584Å, Hα and [SII] at 6717 and 6731Å.

**Table 4.** The ATCA, MOST and PMN imaging properties for the data used in this work.

| Frequency (MHz) | rms Noise (Jy Beam$^{-1}$) | Restoring Beam (arcsec) |
|---|---|---|
| 843 MOST | 0.0030 | 45 |
| 1384 ATCA | 0.0014 | 95 |
| 2432 ATCA | 0.0008 | 58 |
| 4850 PMN | 0.010 | 298 |

## 3.2 SUMSS image of the SNR G332.5-5.6

The Sydney University Molonglo Sky Survey (SUMSS) at 843 MHz (Bock, Large & Sadler, 1999) also completely covered the area of G332.5-5.6. The relevant archival image is shown in Fig. 10. The structure evident at this frequency, and the position and alignment of the three major parallel radio components is very similar to that seen at 4850 MHz (PMN - see below) and in the 13 and 20 cm ATCA images. The more amorphous and fragmented structure of the central component is evident at 843 MHz. We note that the SUMSS image also suffers from missing short spacing antennae and therefore any quantitative study is limited.

## 3.3 PMN image of the SNR G332.5-5.6

In Figure 11 we present the original, lower resolution PMN radio flux data as a combined grey-scale and contour map. These data present the SNR with an overall diameter of ~30'. Three parallel, straight radio filaments, with extent 27' × 10', 22' × 10' and 17.5' × 8', are again seen as for the other radio frequency maps. Based on these data Duncan et al. (1997) concluded that G332.5-5.6, if a bona-fide SNR, does not conform to typical SNR morphology which also explains why it was never incorporated into the Green compilation of known and possible Galactic SNRs.

There are nine PMN putative `point-like' radio sources at 4850 MHz in the immediate vicinity of G332.5-5.6 as recorded in SIMBAD. With the exception of the point sources PMN J1643-5418 and PMN J1642-5436, we consider the remainder to be associated with this SNR (see further discussion in section 3.5). After checking the Kuchar & Clark (1997) all-sky catalogue of flux measurements for HII regions at 4850 MHz, we confirmed that none of these PMN sources is a previously identified HII region.

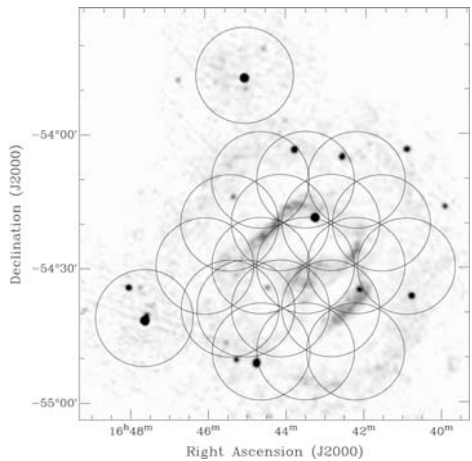

**Figure 7.** Field coverage of the archived ATCA observation at 20/13 cm. Circles represent the ATCA 13-cm primary beams with diameters ~22' at each of 21 pointings. There were 19 pointings which covered the wide field of the SNR together with two extra pointings centered on two strong, point-like sources North and East of the SNR. The grey scale map is a total intensity image of the observed field at 13 cm.

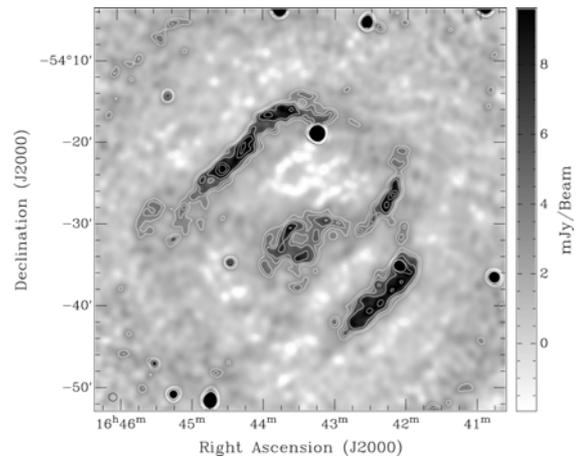

**Figure 8.** ATCA image of G332.5-5.6 at 13 cm. Contour levels are: -3, -1, 3, 4, 5, 7, 10 and 12 mJy beam$^{-1}$. Negative contours are indicated with gray lines. The nominal beam size of 58" is shown to scale in the bottom left corner as a black circle.

The Hα image was compared with the available radio images, noting the positions of the optical filaments relative to the coarser resolution radio filaments and structures. A high level of positional coincidence across much of the remnant is evident. The central broad radio structure, seen at all observed frequencies, is also coincident with significant extended X-ray emission (see later). Weak Hα emission is seen associated with the southern filament and the western edge of the middle component but no Hα emission is seen for the northernmost radio filaments, which is a major structure in all the observed frequencies (see further discussion). It is common that optical filaments and diffuse emissions (mostly in Hα) do not precisely follow radio emission of SNRs (see Green (2006) and optical references therein). Optical spectra from slit positions b), and c), given in Table 1, were obtained close to the outside radio border of G332.5-5.6. At 4850 MHz these radio borders are marked with the lowest positive emission contour of 0.01 Jy beam$^{-1}$ (see Fig. 11).

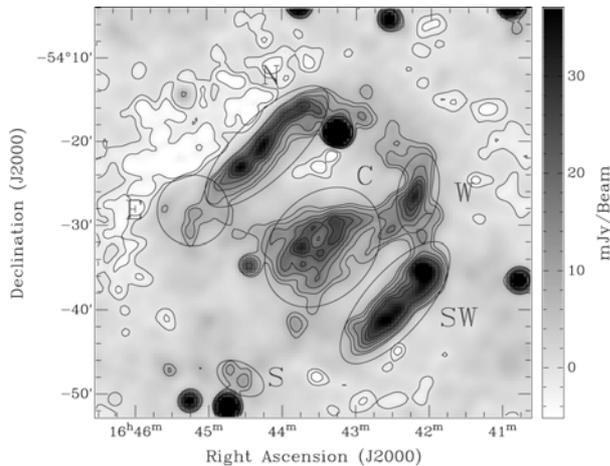

**Figure 9.** ATCA radio continuum mosaic at 20 cm (1384 MHz) of SNR G332.5-5.6. Ellipsoids represent approximate boundaries of the bright semi-distinct SNR radio components N, C, W and SW. The beam size of 9500 is shown in bottom left corner as a black circle.

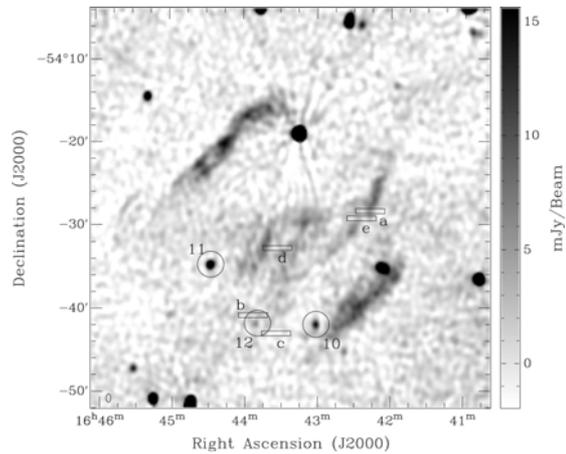

**Figure 10.** 843 MHz SUMSS image of G332.5-5.6. The black rectangles show spectroscopic slit positions with letters corresponding to the positions in Table 1. Three previously uncatalogued point-like radio sources are labeled with black circles numbered 10, 11 and 12. The flux density, in units of mJy beam$^{-1}$, is shown to the right and the beam size of 43" × 53" cosec(|δ|) is shown as a small black ellipse in the bottom left hand corner.

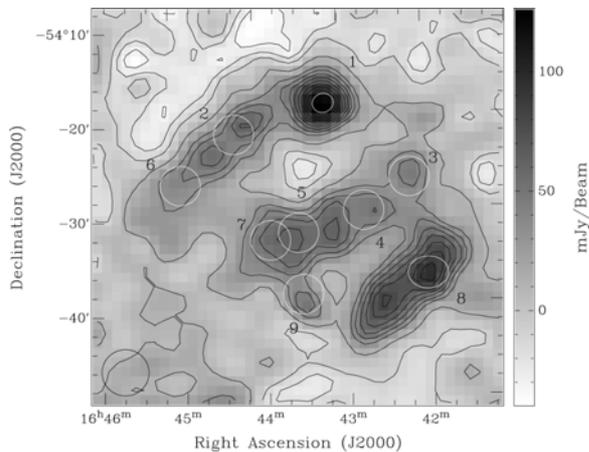

**Figure 11.** PMN image of G332.5-5.6 at 4850 MHz with contour levels -30, -20, -10, 10, 30, 40, 50, 60, 70, 80, 90 and 100 mJy beam$^{-1}$. A flux level grey scale is given at the right. The PMN catalogued sources are indicated with white circles. Sources 1 and 8 have been plotted with ellipses of the quoted sizes as in the PMN catalogue (2.2' × 2.0' and 3.4' × 2.0' respectively). The remaining sources are plotted with 4.2' × 4.2' circles, corresponding to the resolution of the PMN survey. The beam size is shown in the bottom left corner as a black circle.

**Table 5**. Measured radio properties for the G332.5-5.6 sub-components. The components are shown in Fig. 9.

| Filament | Approximate size in arcminutes | Spectral index $\alpha \pm \Delta\alpha$ |
|---|---|---|
| N | 18 × 5 | −0.6 ± 0.1 |
| C | 13 × 11 | −0.8 ± 0.3 |
| W | 7 × 4 | −0.2 ± 0.2 |
| SW | 15 × 5 | −0.5 ± 0.1 |
| Total | 30 × 30 | −0.6 ± 0.1 |

## 3.4 Radio-continuum spectral study

Determining flux densities from the PMN and SUMMS radio data is problematic due to known problems of underestimating flux for such extended sources (for details on the problem with extended objects in the PMN survey refer to Condon, Griffith & Wright 1993; for the SUMSS, Bock, Large & Sadler 1999). However, it is known from other studies (particularly in the SMC/LMC) that the ATCA can recover a significant amount of extended flux given appropriate *uv* coverage - even when mosaicing is used (e.g. see the very thorough and pioneering study of Stanimirovic et al. 1999 and references therein).

Spectral indices were therefore determined with the standard differential spectral index plots method (or T-T method, Costain 1960; Turtle et al. 1962) but using the newly processed ATCA 13 and 20 cm observations. The advantage of this method is that the result is minimally affected by the zero-level errors in the data. Images at both frequencies were constructed from the overlapping *uv* region (between 0.12 and 2.2 k$\lambda$) and convolved with a circular restoring beam of 120 arcsec. The T-T method was separately applied to the four identified bright, sub-components (N, C, W and SW) as shown in Fig. 9 with the results presented in Fig. 12.

The best fit to the data is determined from linear plots and the spectral indices were calculated from the slope of the fitted lines. In order to estimate the error in the slope the regression was done twice, with each image being alternately the abscissa. The derived spectral indices are listed in Table 5 (Col. 3). The large scatter in the T-T plots implies a likely mixture of thermal and non-thermal components. This scatter is especially prominent in the central part of the SNR (component C, top right panel of Fig. 12). The overall negative spectral index for G332.5-5.6 of $\alpha = -0.6 \pm 0.1$ (Fig. 13) is highly typical of SNRs and non-thermal sources.

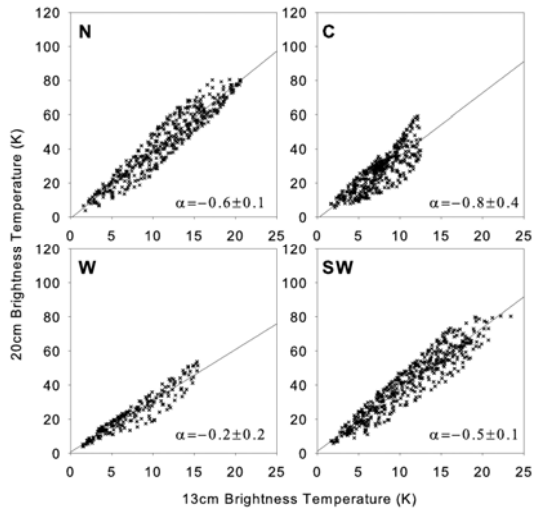
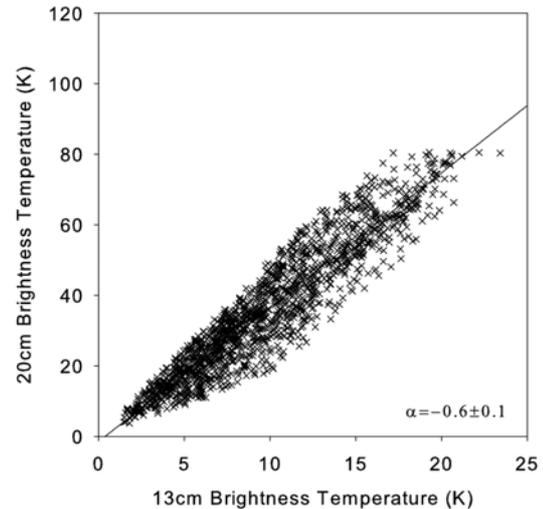

**Figure 12**. T-T plots of the four SNRs bright filaments with the best fitting straight lines for the brightness temperature at 13 cm as abscissa. Spectral indices are calculated from the mean value of fitted line slopes for both frequencies as an abscissa.

**Figure 13.** Combined T-T plot of all 4 G332.5-5.6 bright filaments with the best fitting straight line for the brightness temperature at 13 cm as abscissa. The resultant spectral index of -0.6 is calculated from the mean value of fitted line slopes for both frequencies as an abscissa.

## 3.5 Polarisation measurements of SNR G332.5-5.6

A relatively strong component of linearly polarized emission was detected in the ATCA 13 cm observations (Fig. 14). The average fractional polarization for the bright components N, C and SW is $P_N = 0.42 \pm 0.23$, $P_C = 0.56 \pm 0.30$ and $P_{SW} = 0.37 \pm 0.22$, respectively (where $P_X = \frac{\sqrt{Q_X^2 + U_X^2}}{I_X}$.

The large errors are due to the uncertain polarization fractions at the periphery of the polarization `patches' which varies between 50% and 80%. Usually SNR polarization levels are ~10%, e.g. Milne (1990), Milne, Caswell & Haynes (1993). This derived large fractional polarization may represent only an upper level of the real value. No significant polarization was detected from any of the compact sources found in the observed field.

If this large polarization is real, then together with the more amorphous morphology of the central component compared with the more filamentary appearance of the N and S filaments, this could be an indicator of a pulsar-wind driven nebulae (PWN) in this part of the SNR. A counter argument is the steeper radio spectrum observed in the central region compared to the values derived for the N, W and SW bright components, though the errors are large. The indicative thermal nature seen in the X-ray data (refer section 4) and the lack of a detected pulsar (see below) likewise argues against the PWN hypothesis. Only higher resolution radio and X-ray observations will reveal the true nature of the emission mechanism in the central part of this new SNR.

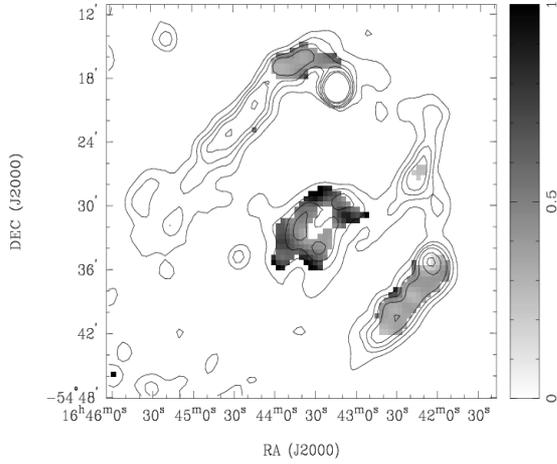 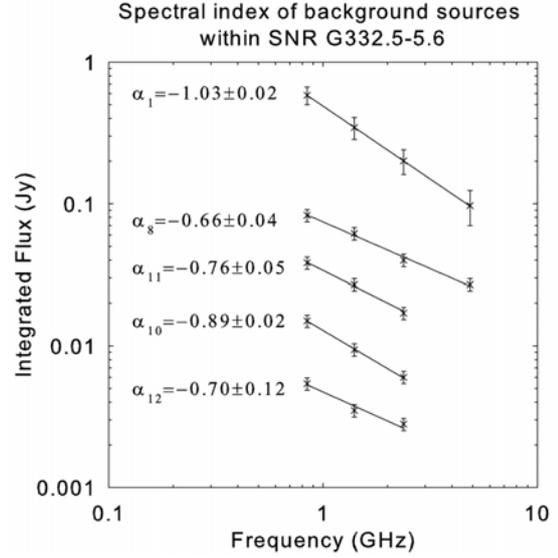

**Figure 14.** Grayscale map of linearly polarized emission from SNR G332.5-5.6 at 13 cm. This image represents the fractional polarization P $(P = \frac{\sqrt{Q^2 + U^2}}{I})$ across the SNR. The overlaid contours are from the ATCA 13 cm image at 2, 6, 10, 15, 25 and 35 mJy beam$^{-1}$.

**Figure 15.** The radio spectrum of five background point sources in the SNR G332.5-5.6 field. The power law regression lines were constructed from the measured integrated flux densities.

**Table 6.** Catalogued radio sources in the field of SNR G332.5-5.6. We include three previously uncatalogued sources. We present flux density measurement only for the five point sources projected onto G332.5-5.6. The positions for these five sources where obtained from our 13-cm image (with the 6$^{th}$ antenna) where the best resolution can be achieved.

| Source No[†] | Catalogue Name | RA (J2000) h m s | Dec (J2000) ° ′ ″ | $S_{843\,MHz}$ (mJy) | $S_{1384\,MHz}$ (mJy) | $S_{2432\,MHz}$ (mJy) | $S_{4850\,MHz}$ (mJy) | Spectral Index |
|---|---|---|---|---|---|---|---|---|
| 1 | J1643-5418 | 16 43 14.6 | -54 19 06.9 | 584 ± 61 | 345 ± 40 | 210 ± 20 | 97 ± 10 | −1.03 ± 0.02 |
| 2 | J1644-5421 | 16 44 17.0 | -54 21 23 | | | | | |
| 3 | J1642-5426 | 16 42 12.4 | -54 26 38 | | | | | |
| 4 | J1642-5429 | 16 42 46.0 | -54 29 59 | | | | | |
| 5 | J1643-5432a | 16 43 33.7 | -54 32 06 | | | | | |
| 6 | J1644-5426 | 16 44 57.2 | -54 26 30 | | | | | |
| 7 | J1643-5432b | 16 43 54.2 | -54 32 46 | | | | | |
| 8 | J1642-5436 | 16 42 05.6 | -54 35 16.1 | 83 ± 9 | 61 ± 7 | 40 ± 4 | 27 ± 3 | −0.66 ± 0.04 |
| 9 | J1643-5438 | 16 43 32.2 | -54 38 39 | | | | | |
| 10 | | 16 43 01.2 | -54 42 05.9 | 15 ± 2 | 9.4 ± 1.0 | 6.0 ± 1.0 | | −0.89 ± 0.02 |
| 11 | | 16 44 28.0 | -54 34 52.6 | 38 ± 4 | 27 ± 3 | 17 ± 2 | | −0.76 ± 0.05 |
| 12 | | 16 43 52.0 | -54 42 03.6 | 3.5 ± 0.4 | 3.5 ± 1.0 | 2.8 ± 1.0 | | −0.70 ± 0.12 |

[†] Sources 1-9 marked in Figure 11, sources 10-12 marked in Figure 10

### 3.6 Background sources in the field of SNR G332.5-5.6

The 9 sources found in the PMN survey within a 1° field of the SNR are plotted in Fig. 11 and listed in Table 6. Sizes are given in the figure only for two resolved sources: PMN J1642-5436 and PMN J1643-5418. The rest are unresolved and are plotted with 4.2' diameter circles equal to the PMN beam size.

Three new point sources detected by ATCA and MOST but not seen in the PMN survey are also listed and marked as 10, 11 and 12 in Fig. 10). For the five compact radio sources in the vicinity of the SNR (Sources No. 1, 8, 10, 11 and 12 Figs. 10 and 11) integrated flux densities were measured from the PMN, SUMSS and ATCA images. For four of them (1, 10, 11 and 12) the MIRIAD task IMFIT was used. For source 8 (PMN J1642-5436) embedded into the SW filament on Fig. 11, a tight region around the source was constructed and the MIRIAD task IMSTAT used to estimate its integrated flux density. The spec-tral indices found for these sources from power-law regression slope fits (Fig. 15) were all negative, ranging between α = -1.03±0.02 and α = -0.66 ± 0.04. Results, together with the derived integrated flux densities, are presented in Table 6 (Col. 9).

Only the background object source 8 (PMN J1642-5436; Table 6) could be resolved using *uv* data from all 6 ATCA antennas at 20 cm (Fig. 16). At high resolution the morphology is revealed as a central point-like source with adjacent opposing jet-like structures. It has a non-thermal spectral index of α = -0.66 ± 0.04 which is typical of an AGN/double-lobed radio galaxy.

It is very likely that the majority of the listed PMN sources in Table 6 (7 out of 9) are actually part of SNR G332.5-5.6. Source 3 (PMN J1642-5426) coincides directly with the paperclip. There is a point source at the position of source 8 (PMN J1642-5436) but the size quoted for this source in the catalogue shows that it could not be disentangled from the emission in the S-W filament. The bright source 1 (PMN J1643-5418), although recognized as distinct, is positionally biased towards the N-E filament. For these latter two sources, the position in Table 6 has been recalculated based on our newly processed images. For other PMN sources, the coordinates are given as quoted in the PMN catalogue.

## 4 ROSAT X-RAY OBSERVATIONS OF SNR G332.5-5.6

Motivated by the fact that we have new optical and radio confirmation for this Galactic SNR, we searched for any possible associated X-ray sources or a pulsar. No pulsars are known in the vicinity of G332.5-5.6 (Manchester et al. 2005[2]) but a SIMBAD search confirmed three X-ray sources inside the overall 300 PMN contour for the SNR. The NE ¯lament matches the small ROSAT X-ray source RX J164437.6-542120 while the SW filament coincides with RX J164231.3-544242. The third and most prominent X-ray source RX J164251.3-542955, is extended and completely matches the central broadened radio component of G332.5-5.6.

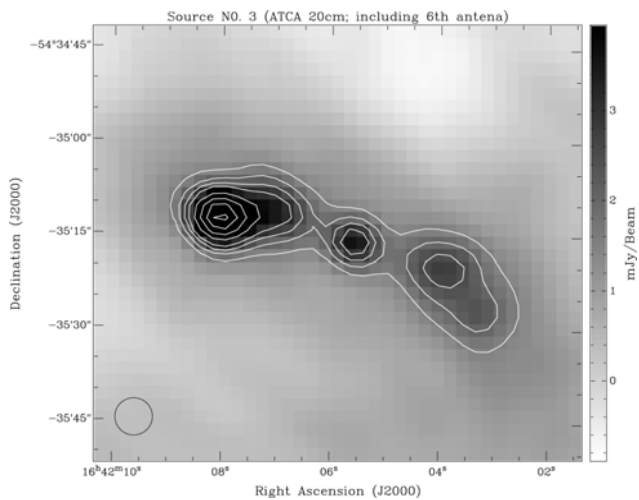
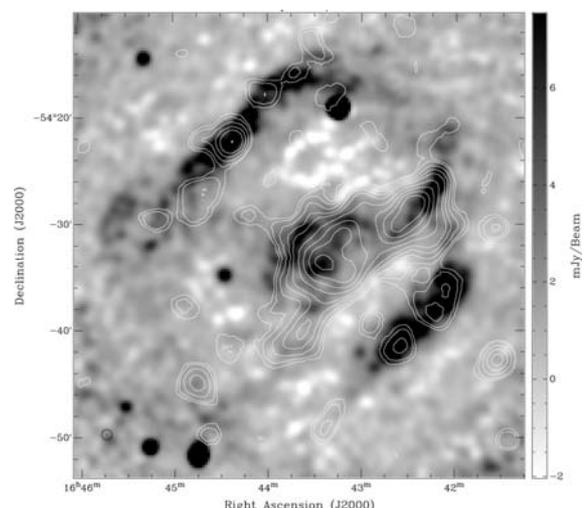

**Figure 16**. ATCA image of the newly resolved `double-lobed' PMN source J1642-5436 (source 8; Fig. 11) at 20 cm. Contours are: 1.5, 2, 2.5, 3, 4, 5, 6 and 7 mJy beam$^{-1}$. The beam size of 6 arcsec diameter is shown in the bottom left corner. This object is probably a background radio galaxy/AGN.

**Figure 17.** ROSAT X-ray image (0.5-2.4 keV) of G332.5-5.6 (contours: 0.12, 0.15, 0.175, 0.2, 0.25, 0.3, 0.35, 0.4 and 0.45 ×10$^{-4}$cts sec$^{-1}$) overlaid on the ATCA 13-cm gray-scale image. The X-ray emission, which is predominantly of thermal origin, nicely matches several radio features.

The ROSAT data base[3] revealed that a field including G332.5-5.6 was observed with the ROSAT PSPC (Trümper 1982) for 3043 seconds on 8$^{th}$ September 1993. The archived data was used to create new, cleaned, exposure-corrected and smoothed X-ray images. These were then combined into broad (0.1-2.4 keV), soft (0.1-0.4 keV), hard (0.5-2.0 keV), hard 1 (0.5-0.9 keV) and hard 2 (0.9-2.0 keV) X-ray band images. Figure 17 shows the resultant hard-band (0.5-2.0 keV) image as contours overlaid on the 13-cm radio map for a 40' region centred on the SNR. The existing PSPC observations were binned to 30 arcsec pixels and smoothed (20) for better representation. The X- ray emission is seen to match the prominent radio features of G332.5-5.6 with both the central elongated clumpy radio structure and the outer parallel features in the NE and SW regions also detected in X-rays.

An X-ray spectral analysis of the extended region around RA = 16$^h$43$^m$30$^s$ and δ = -54°33' (J2000) out to a radius of 28.5' radius was undertaken. At each 0.7125' incremental distance from the centre of the source the X-ray flux was summed in an annulus (ring) at that distance and, after exposure and background correction, the resulting average value used to create a radial profile in the standard manner, e.g. Mavromatakis et al. (2004). The resultant profile in Fig. 18 averages out the most obvious clumpy structure but also takes care of the low level but extended brightness regions. It shows a steep decline in the log of X-ray counts out to about 14'. Thereafter there is an indication of a shell, which for the centre position chosen for the profile, is coincident with the position of the north-eastern radio/X-ray filament. The shape of the core of the radial profile as well as the excess of counts at the shell position depend of course on the choice of the centre position. This is not unique because of the asymmetry of the shell filaments and the two, resolved patches in the interior of the remnant. By varying the centre position for the radial profile a minimum radial extent of 17.5' for the remnant is obtained, before the profile steps down to the background level.

---

[2] see also http://www.atnf.csiro.au/research/pulsar/psrcat
[3] http://www.xray.mpe.mpg.de/cgi-bin/rosat/seq-browser

It is possible that G332.5-5.6 is another example of the rare class of mixed morphology SNRs (Rho & Petre 1998), i.e. a bright centre and a diffuse but filled body with a low brightness shell (if any) in X-rays. In the radio and the optical there are bright rims, i.e. a fractured shell, but the diffuse interior is absent.

Most of the X-ray profile follows an exponential decline and is therefore very similar to the SNR G82.2+5.3 (Mavromatakis et al. 2004; also classified as mixed morphology; see Fig. 6 of that paper). At the moment, there are few explanations for how these remnants arise (Rho & Petre 1998; Lazendic & Slane 2006). We have also looked for spectral variations across different cuts in the X-ray data but no significant differences were found.

A Raymond-Smith thermal spectrum gives a satisfactory fit to the data giving a temperature of ~ 0:25 keV (comparable with most regions of the Vela SNR) and the interstellar absorption is $0.83 \times 10^{21}$ per cm$^2$. In contrast, the total Galactic absorption is 4 times higher. Hence the object is clearly in our Galaxy and is probably relatively close-by and, because of the large observed angular extent, dynamically old. The high Galactic latitude of 5.5° for this SNR is puzzling. At the distance of ~ 3-4 kpc as suggested from the various distance estimates given later in Section 5 it would be at ~300-400 pc above the plane and have a physical diameter of ~ 30 pc.

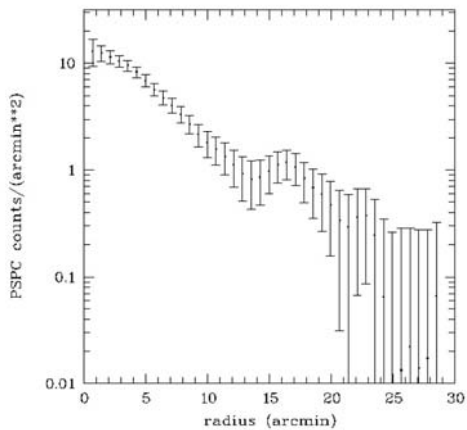

**Figure 18.** ROSAT X-ray profile of SNR G332.5-5.6. as azimuthally measured from the central X-ray peak zone. The apparent full X-ray radial extent of this SNR amounts to ~ 23'. The local peak at ~ 17' is coincident with the position of the north eastern radio/X-ray filament.

**5 THE DISTANCE TO G332.5-5.6**

Establishing a reliable distance to an SNR is difficult, especially from radio observations. This was the motive for us to try to determine distance estimates from optical observations. Sometimes it is possible to obtain a kinematic distance to a nebula if a proper system velocity from various measures across the expanding shell can be determined and then used in a Galactic rotation curve model. However we cannot be sure that the radial velocities, as derived from the various optical filaments (refer Table 3), are representative of the full nebula (due to the effects of variable extinction), especially given the comparison between the optical and radio distributions of the remnant.

The Hβ/Hα flux ratio of the brightest optical filament implies modest foreground extinction is present at this location. The average reddening E(B - V) = 0.27 is not significantly different to the asymptotic reddening, E(B - V) = 0.35 through the Galactic disk in this direction, as calculated by Schlegel, Finkbeiner & Davis (1998), making application of the extinction-distance method problematic. The Galactic latitude of 5.6° is high for an SNR (only ~4% of the SNRs in Green's catalogue have greater values of |b|). The remnant is considered though, on the basis of the rich shock-excited line spectrum, to be within the dust layer in the Galactic disk. Assuming a conservative height for the dust layer of 300 pc, a tentative upper limit for the SNR's distance of 3.1 kpc can be derived.

Though the use of Σ-D (or surface brightness to diameter) relation (Shklovsky 1960) for Galactic SNRs distance is controversial (e.g. Case & Bhattacharya (1998), Huang & Thaddeus (1985)) we used this as another basis for a distance estimate as described below. Despite all the uncertainties of this method, e.g. Arbutina & Urošević (2005), it was shown by Stupar et al. (2007), for the PMN sample, that if SNRs are chosen according to rigid criteria, then the Σ-D relationship can provide useful Galactic SNR distance estimates. Arbutina & Urošević (2005) also found a distance solution for remnants in low-density interstellar media. If we assume that G332.5-5.6 is evolving in such a low density environment because of its height above the Galactic plane (|b| of 5.6°) then their Σ-D solution can be used to estimate a distance for G332.5-5.6. However, it is also true that the observed Balmer emission must arise in a high density environment. Obviously, the local area of the SNR must be within a higher density environment otherwise we would not detect Balmer emission and other optical emission lines.

Accepting that the radio spectral index between 1384 and 2432 MHz is α = -0.6 ± 0.1, then the surface brightness at 1 GHz is $4.4 \times 10^{-22}$ W m$^{-2}$Hz$^{-1}$sr$^{-1}$. Applying this value and taking a diameter of ~ 30' to the Arbutina & Urošević (2005) Σ-D solution for low-density environments, yields a distance $d \sim 4$ kpc.

Finally, these crude results were compared with the recently determined independent distance estimate made for this object by Reynoso & Green (2007) on the basis of the HI λ21-cm line seen in absorption against the continuum emission. They adopt -45 km s$^{-1}$ as the SNR system velocity and using $R_\odot$ = 8.5 kpc and $\Theta_\odot$ = 220 km s$^{-1}$, determined a distance $d$ = 3.4 kpc.

This is in close agreement to the upper limit derived from assumption of the SNR's location within the dust layer and in good agreement with the Σ-D value. A distance of 3 - 4 kpc for this SNR therefore seems likely.

**6 DISCUSSION**

As SNRs are non-thermal they are generally strong radio emitters. Optical emission is often seen if the foreground extinction is not too severe. There can be coexistence between Hα, X-ray and radio continuum emission for remnants (Cram, Green & Bock 1998). In the case where there is propagation of the supernova blast into a hot, low-density ISM little Hα emission can be seen. However, propagation into a cool, dense ISM will lead to Balmer emission, as for G332.5-5.6. The observed values and stability of line ratios for the different optical components, and the observed 10% stability of the density-sensitive [SII] 6717/6731 ratio are in agreement with the optical diagnostic line ratios of Fesen, Blair & Kirshner (1985) for identifying SNRs. The high ratios of [SII]/Hα indicates that the optical emission originates from shock heated gas. The observed [SII] 6717/6731 ratios range between 1 and 1.4 (~ 1.23 on average) which indicates a low electron density in the filaments (Osterbrock & Ferland 2006). Using the IRAF nebular package (Shaw & Dufour 1995) we estimated that the average electron density from the spectra taken at all six slit positions is ~ 240 $cm^{-3}$ (ranging from 40 to 600 $cm^{-3}$) from the [SII] 6717/6731 ratio and assuming a nebula temperature of 10 000 K. The first set of spectroscopic observations from the good photometric night (February 2004, see Table 1) was used to estimate the logarithmic extinction at Hβ as $c$ = 0.39 using Seaton (1979), which is equivalent to a reddening of E(B - V) = 0.27.

The morphological structure of G332.5-5.6 was examined and the positions of the filaments as seen separately in the respective optical (Hα), radio and X-ray images compared. The newly detected optical line-emission has an excellent positional and position-angle match with the central radio structure. There is also an excellent match between the radio (in all frequencies) and X-ray images, again, especially for the central, brightest oval component. Two of the major optical filaments, positions b) and c), Fig. 10, appear to extend over the radio filament borders, but these two optical filaments are clearly within the outer boundaries of the X-ray emission. This is in line with the observed spectral lines being created by shock motion inside the SNR, and that the radio brightness is possibly underestimated.

A comparison of the positional coincidence of optical and radio filaments for Galactic SNRs gives two basic groups. The first is when optical emission (mostly in Hα and [SII]) is very strong and follows, if not the whole remnant, then at least one major component of the radio emission, usually in the form of an arc. An example of this SNR type can be seen in the images of CTB 1 in the light of Hα (Fesen et al. 1997) and in the radio (Dickel & Willis 1980). The other group, of which we consider G332.5-5.6 a member, are composed of (mostly old) remnants where the optical emission and fine filaments are fragmented compared with the more coherent radio images. Furthermore these filaments do not closely follow the radio emission. This optical fragmentation in some cases crosses over the radio borders. Actually, we cannot say with certainty that filaments b) and c) are out of the remnants radio borders as some larger structure around G332.5-5.6 may be missing from the presented radio maps as they suffer from apparent negative flux levels. This is due to the lack of zero spacing flux and incomplete *uv* coverage so that interpretation of the radio maps must be undertaken with this in mind. Finally, it should be noted that localised variable extinction can weaken or eliminate the optical detection of the SNR across parts of the remnant.

A similarity in the overall optical filamentary structure has been noticed between this remnant, G332.5-5.6, and remnant G291.0-0.1 (Whiteoak & Green 1996). Both these remnants comprise three parallel filaments (the only difference is in position angle) with a bright central component seen in X-rays. G291.0-0.1 is classified as a mixed morphology Galactic SNR (Wilson 1986) with most probably two shells surrounding strong central emission. However, in the radio, the central part and two possible side shells of G332.5-5.6 are similar in brightness. This is different to the situation for G291.0-0.1. This does not prevent us classifying G332.5-5.6 as a mixed morphology remnant.

Further systematic investigation, and in particular additional radio frequency observations, are required to show if the radio spectral index varies and is locally flatter in the central portion (with values that would be expected to be α=-0.1 to -0.3, unlike the -0.6 currently observed for component C, Fig 9), which is typical for plerions (Weiler & Sramek 1998). If a plerion is present then G332.5-5.6 (and most probably G291.0-0.1) would evolve into a composite and then into a shell remnant (Lozinskaya 1981).

Assuming a distance of ~ 3 - 4 kpc (as implied by the various distance indicators used), the implied physical diameter is ~30 pc. This suggests that G332.5-5.6 is a rather old or middle-aged SNR. The radio-continuum images strongly support this with a central fragmented structure typical for SNRs in the middle to late evolutionary stage, a conclusion which also follows from the X-ray analysis.

In conclusion we have brought together many lines of evidence from multi-wavelength data which presents a compelling picture of the veracity of this new and interesting SNR. Firstly the integrated radio spectral index α=-0.6 ± 0.1, confirms the non-thermal nature of the source. Secondly our new optical imagery in the Hα emission line provides strong morphological clues while the observed structure is shown to match in orientation and extent with several radio components of the SNR. Thirdly, our optical spectroscopy reveals highly indicative SNR spectral signatures, especially the strong [SII] relative to Hα which betrays the presence of shock heated gas. Finally, we have the corroborating X-ray observations. We therefore conclude that G332.5-5.6 is a bona-fide new Galactic SNR. More detailed X-ray investigations (XMM and Chandra) could give a definitive answer on the fuller character of G332.5-5.6.

**ACKNOWLEDGEMENTS**

We are thankful to the staff of the Wide Field Astronomy Unit at the Royal Observatory Edinburgh for making the AAO/UKST Hα Survey of the Southern Galactic Plane available online. QAP, MS and DJF thank ANSTO for travel grants that enabled the spectroscopic observations to be undertaken. The ATCA is part of the Australia Telescope, which is funded by the Commonwealth of Australia for operation as a National Facility managed by CSIRO. This research has made use of the SIMBAD database, operated at CDS, Strasbourg, France. We would also like to express our sincere thanks to the anonymous referee for some excellent suggestions which have significantly improved this paper.